\documentclass[reprint,amsmath,amssymb,aps,prl,superscriptaddress]{revtex4-2} 

\newcommand{\Deltadon}{\Delta^\mathrm{Don}}

\newcommand{\cs}{c_\mathrm{salt}}
\newcommand{\cmon}{c_\mathrm{mon}}
\newcommand{\zmon}{z_\mathrm{mon}}

\newcommand{\cref}{c^{\ominus}}

\newcommand{\pKa}{\mathrm{p}K_{\mathrm{A}}}
\newcommand{\pKeff}{\pKa^{\mathrm{eff}}}
\newcommand{\app}{^{\mathrm{app}}}
\newcommand{\pH}{\mathrm{pH}}

\usepackage[normalem]{ulem} 
\newcommand{\soutr}[1]{\sout{\textcolor{red}{#1}}}
\newcommand{\soutb}[1]{\sout{\textcolor{blue}{#1}}}

\renewcommand{\soutr}[1]{\relax}
\renewcommand{\soutb}[1]{\relax}  

\usepackage{graphicx}
\usepackage[hidelinks]{hyperref}
\usepackage[usenames,dvipsnames]{color}
\usepackage[version=3]{mhchem}

\makeatletter
\let\old@makecaption=\@makecaption
\usepackage{subcaption}
\let\@makecaption=\old@makecaption
\makeatother

\usepackage{xr}
\begin{document}

\preprint{APS/123-QED}

\title{Explaining Giant Apparent $\pKa$ Shifts in Weak Polyelectrolyte Brushes}

\author{David Beyer}
\affiliation{Institute for Computational Physics, University of Stuttgart, D-70569 Stuttgart, Germany}

\author{Peter Košovan}
\email{peter.kosovan@natur.cuni.cz}
\affiliation{Department of Physical and Macromolecular Chemistry, Charles University, Prague, Czechia}

\author{Christian Holm}%
\email{holm@icp.uni-stuttgart.de}
\affiliation{Institute for Computational Physics, University of Stuttgart, D-70569 Stuttgart, Germany}

\date{\today}

\begin{abstract}
Recent experiments on weak polyelectrolyte brushes found marked shifts in the effective $\pKa$ that are linear in the logarithm of the salt concentration.
Comparing explicit-particle simulations with mean-field calculations we show that for high grafting densities the salt concentration effect can be explained using the ideal Donnan theory, but for low grafting densities the full shift is due to a combination of the Donnan effect and the polyelectrolyte effect. 
The latter originates from electrostatic correlations which are neglected in the Donnan picture and which are only approximately included in the mean-field theory. 
Moreover, we demonstrate that the magnitude of the polyelectrolyte effect is almost invariant with respect to salt concentration but depends on the grafting density of the brush. 
This invariance is due to a complex cancellation of multiple effects.
Based on our results, we show how the experimentally determined $\pKa$ shifts may be used to infer the grafting density of brushes, a parameter that is difficult to measure directly.
\end{abstract}

\maketitle

{\em Introduction.} 
Recent experiments by Ferrand-Drake del Castillo et al. \cite{ferranddrake20a} have demonstrated marked shifts in the effective $\pKa$ of weak polyelectrolyte brushes, tunable by varying the salt concentration.
Furthermore, they used three different polymers, including both acidic and basic polyelectrolytes (PEs), to show that these shifts are approximately linear in the logarithm of the salt concentration.
In a related study \cite{ferranddrake20b}, they demonstrated that the tunable response of these brushes to variations in pH and salt concentration makes them excellent candidates for high-capacity protein capture and release through pH adjustment.

Similar $\pKa$ shifts have been predicted by mean-field models and could qualitatively be explained using the ideal Donnan theory \cite{nap06a,nap17a,borisov11a,zhulina17a,zhulina11a,zhulina12a}.
In the Donnan theory, the polyelectrolyte brush is approximated as a homogeneous phase in equilibrium with the bulk solution, which acts as an infinite reservoir of ions.
By assuming that the brush confines all of its counterions, a Donnan potential emerges between the brush and the bulk solution.
If the concentration of the polymer-bound charges is much higher than the ionic strength of the solution, $I$, the Donnan potential can be approximated as
\begin{align}
\psi^\mathrm{Don} \approx \frac{k_\mathrm{B}T}{\zmon e}\ln\left(\frac{\alpha\cmon}{I}\right).
\label{eq:Donnan_potential}
\end{align}
Here, $\alpha$ denotes the degree of ionization of the brush, $\cmon$ is the total concentration of monomeric units inside the brush, and $\zmon=\pm 1$ stands for the valency in the ionized state.
If we define the effective $\pKa$ as the pH value at which the polymer is 50\% ionized, we can express its shift as
\begin{align}
\pKeff \equiv \pH \left(\alpha=0.5\right) = \pKa + \Delta\stackrel{\mathrm{ideal}}{=} \pKa + \Delta^\mathrm{Don}.
\end{align}
In an ideal system, the shift $\Delta$ depends only on the Donnan contribution $\Delta^\mathrm{Don}$.
However, in general, also electrostatic interactions contribute to the shift.
Using \autoref{eq:Donnan_potential}, the Donnan contribution can be expressed as
\begin{align}
\begin{split}
\Deltadon &
    \approx -\zmon\log_{10}\left(\frac{\cmon}{2I}\right)\\ 
    &= \zmon \biggl[ \log_{10}\left(\frac{I}{\cref}\right) - \log_{10}\left(\frac{\cmon}{2\cref}\right)\biggr],
\label{eq:Donnan-variation}
\end{split}
\end{align}
where $\cref$ is an arbitrary reference concentration, usually chosen as $\cref=1\,\mathrm{M}$.
If the concentration of ionized monomers at $\alpha=0.5$ does not significantly change with the ionic strength, then the second term in \autoref{eq:Donnan-variation} is constant, and $\Deltadon$ scales linearly with the logarithm of the ionic strength, as confirmed by recent experimental measurements of the Donnan potential in polyelectrolyte membranes \cite{gokturk22a}.
Considering these findings, the experiments by Ferrand-Drake del Castillo et al. \cite{ferranddrake20a} seem to confirm that the Donnan approximation can be used to describe $\pKa$ shifts in polyelectrolyte brushes. 
However, in this letter, we show otherwise. 
Our results demonstrate that the Donnan approximation is not necessarily valid and that fully explaining the effect of ionic strength on $\pKa$ requires going beyond this approximation.

Numerical mean-field models have been introduced to alleviate some approximations used in the Donnan theory \cite{nap06a,nap14a,nap17a,zhulina11a,zhulina17a,nap17a,leonforte16a,israels94a,israels94b}.
Within the mean-field approximation, particle-particle interactions are replaced by interactions with an average field, proportional to the mean density at a specific location.
When applied to polyelectrolyte brushes, these models explicitly account for density variations perpendicular to the surface, while averaging over density variations parallel to the surface.
These models also account for electrostatic interactions on the mean-field level by solving the Poisson-Boltzmann equation.
Lastly, local density fields enable us to calculate local variations in ionization states as well.
The density profiles obtained from the mean-field calculations provide a more refined description of the brush-solution interface than the ideal Donnan theory.
Nevertheless, the mean-field approximation works well for polyelectrolyte systems only when inter-chain interactions prevail over intra-chain interactions \cite{uhlik14a}; otherwise, the applicability of the mean-field theory is limited.

Based on the mean-field picture, we can distinguish brushes in an osmotic or salted regime \cite{zhulina17a, israels94a, ballauff06b}.
The osmotic regime occurs in a densely grafted brush when the charge density inside the brush is much higher than ionic strength of the bulk solution.
Consequently, electrostatic interactions are screened, and the brush properties are controlled by the osmotic pressure of counterions and Donnan partitioning.
Characteristic of the osmotic brush regime is that the swelling of the brush is essentially independent of the ionic strength in the solution and the grafting density.
As the bulk ionic strength becomes comparable to the charge density inside the brush, the brush transitions into the salted brush regime.
In this regime, the swelling of the brush is predicted to be dependent on both the bulk ionic strength and the grafting density.

\begin{figure}
\includegraphics[width=0.45\textwidth]{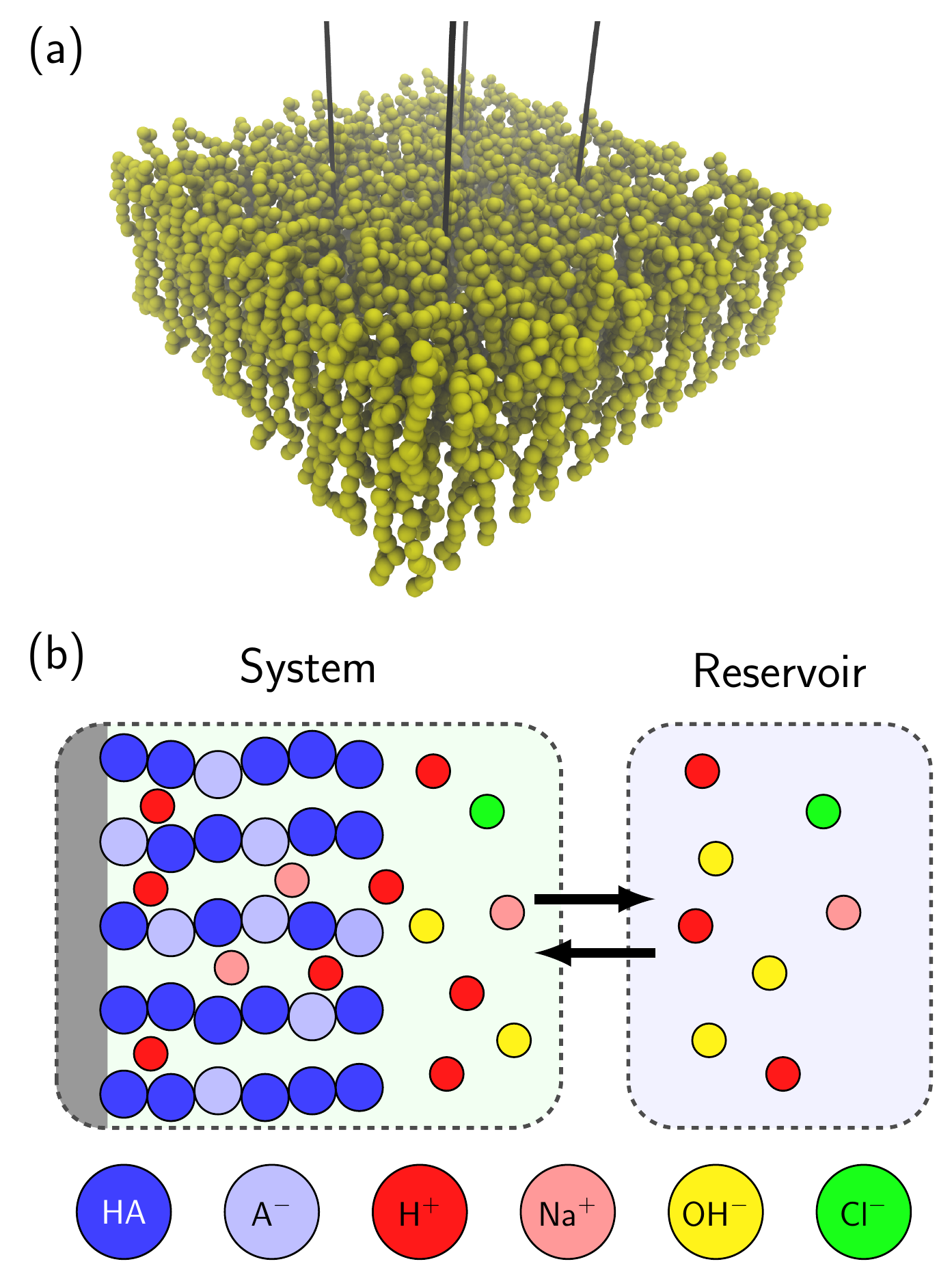}
\caption{\label{fig:snapshot_brush}
(a): Snapshot of the brush model (small ions are not shown for clarity).
Black lines bound to the simulation box and additional periodic images in the $x$-$y$-plane are shown to illustrate the periodic boundary conditions.
The snapshot was produced using VMD \cite{humphrey96a}.
(b): Schematic representation of the simulation setup: simulation box containing the brush, coupled to a reservoir at a fixed pH and salt concentration.
}
\end{figure}

{\em Simulation model and method.} 
To go beyond both, Donnan and mean-field approximations, we constructed a coarse-grained, particle-based simulation model of a polyelectrolyte brush, as shown in \autoref{fig:snapshot_brush}.
Our model consisted of $25$ chains, each of which contained 25 monomers.
The first monomer of each chain was fixed to an immobile, purely repulsive flat surface, explicitly modelling the wall to which the brush is grafted.
Furthermore, our model also explicitly accounted for the interface between the brush and the solution.
The explicit interface between the brush and the solution allowed us to determine the equilibrium swelling of the brush for a given reservoir composition using a single simulation.
Because experimental grafting densities were not available \cite{ferranddrake20a}, we performed simulations at two different grafting densities, within a plausible range, $\Gamma=0.79/\mathrm{nm}^2$ and $\Gamma=0.079/\mathrm{nm}^2$.
The polymer chains were modelled using a generic bead-spring model, derived from the Kremer-Grest model \cite{grest86a}.
Small ions (\ce{Na+}, \ce{Cl-}, \ce{H+}, \ce{OH-}) were represented as explicit particles, whereas the solvent effects were only included implicitly through the relative dielectric constant.
The interactions included short-range steric repulsion between all particles, and the full unscreened Coulomb potential between charged particles.
To represent a flat surface, we used 2D-periodic boundary conditions in a slab.
All monomers were treated as weak acids with $\mathrm{p}K_\mathrm{A}=4.0$ as a typical value for acrylic polymers \cite{katchalsky49a}.
The ionization equilibrium of the acidic monomers and the exchange of small ions with the bulk were simulated using the grand-reaction method \cite{landsgesell20b}.
Full technical details of the model and the simulation method are provided in the ESI.
All simulations were performed using the open-source simulation software ESPResSo \cite{weik19a}.

\begin{figure*}
\begin{subfigure}[t]{0.499999\textwidth}
\includegraphics[width=\textwidth]{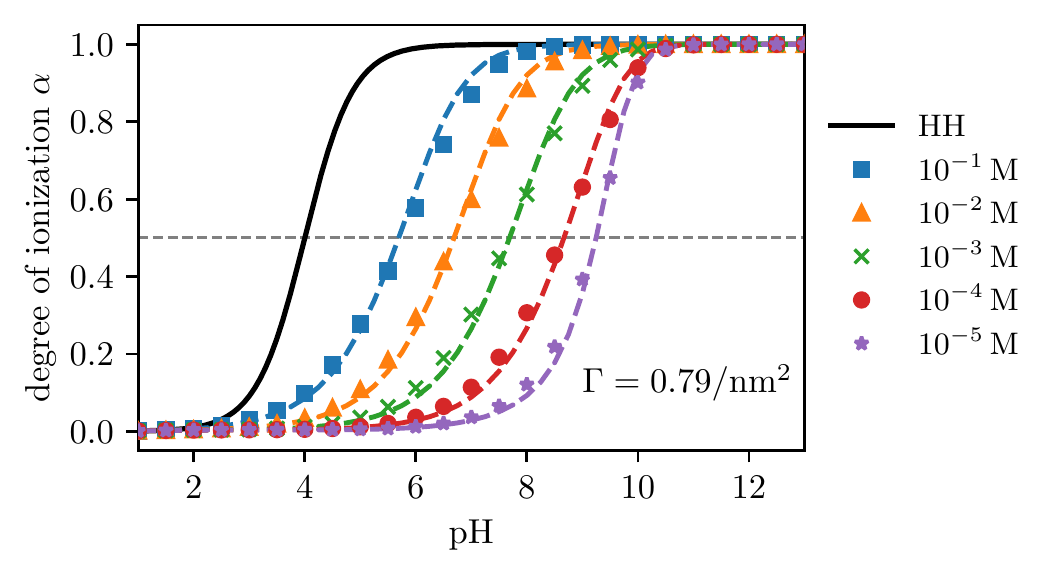}
\caption{}
\end{subfigure}
\begin{subfigure}[t]{0.3999999\textwidth}
\includegraphics[width=\textwidth]{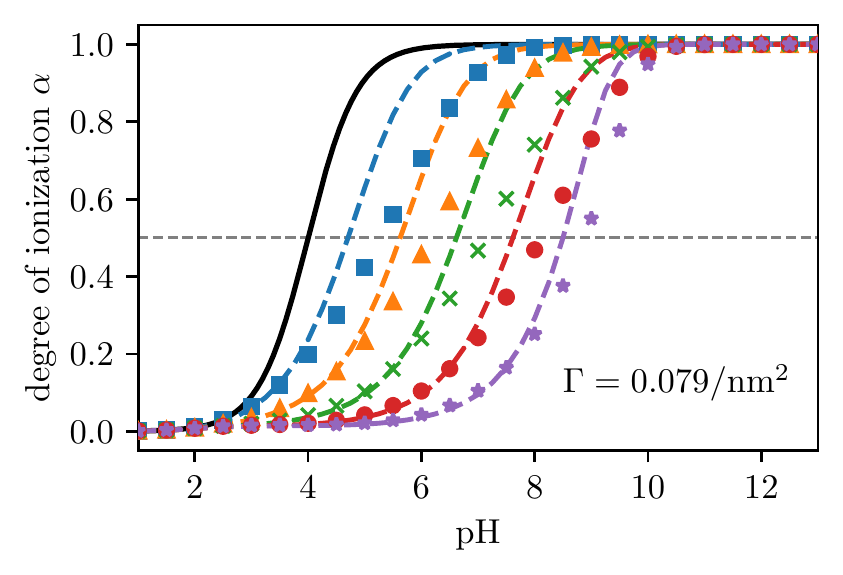}
\caption{}
\end{subfigure}
\caption{\label{fig:alpha_brush}
Titration curve of the weak polyelectrolyte brush with different grafting densities as a function of the pH and for different salt concentrations. 
Markers correspond to simulation results, while the dashed lines indicate the SCF results.
As a reference, the ideal titration curve is shown as described by the Henderson-Hasselbalch equation (HH).
}
\end{figure*}

{\em Results.} 
\autoref{fig:alpha_brush} confirms that the simulation results qualitatively reproduce the experimentally observed $\pKa$ shifts as a function of salt concentration.
Additionally, the simulation results at high grafting density, ($\Gamma=0.79/\mathrm{nm}^2$, \autoref{fig:alpha_brush}a) are quantitatively matched by mean-field calculations for the same brush using the Scheutjens-Fleer self-consistent field (SCF) theory (see ESI for technical details).
Simulation results for the same system at a low grafting density ($\Gamma=0.079/\mathrm{nm}^2$, \autoref{fig:alpha_brush}b) also display qualitatively similar shifts, albeit greater than those obtained from SCF calculations.
This discrepancy between mean-field calculations and simulations indicates that an additional effect contributes to the results, and that this effect is neglected by both, the Donnan theory, and the mean-field approximation.

\begin{figure}
\includegraphics[width=.45\textwidth]{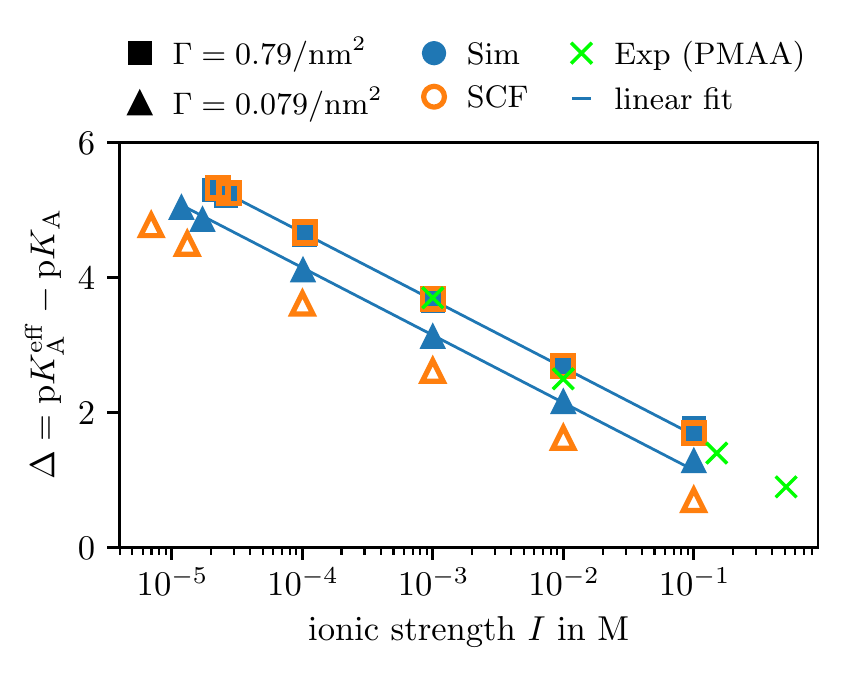}
\caption{\label{fig:effective_pK_brush}
The $\pKa$ shift of brushes with different grafting densities as a function of ionic strength. 
Solid symbols represent data from particle-based simulations, whereas empty symbols represent SCF results. 
}
\end{figure}

\autoref{fig:effective_pK_brush} shows that both datasets, at high and low grafting density, exhibit the same Donnan-like variation of the $\pKa$ shift.
At a high grafting density, simulations and mean-field calculations yield the same $\pKa$ shift.
In addition, they match the $\pKa$ shift of PMAA brushes determined experimentally in Ref. \cite{ferranddrake20a}.
At a low grafting density, the slope of  the $\pKa$ shift as a function of ionic strength is the same in mean-field calculations and simulations, but the absolute shift predicted by the mean-field theory is significantly smaller.
This comparison demonstrates that the Donnan-like variation of the $\pKa$ shift does not imply that the Donnan theory fully describes the problem.

For analyzing the $\pKa$ shift, it is important to recognize that the ionic strength $I=(c_{\ce{Na+}}+c_{\ce{H+}}+c_{\ce{Cl-}}+c_{\ce{OH-}})/2$ does not have the same value as the salt concentration $\cs$.
While they differ only negligibly at high salt concentrations, at low salt concentrations and extreme pH-values, the ionic strength is higher than the salt concentration due to the significant contribution to screening of \ce{H+} or \ce{OH-} ions.
Therefore, plotting the $\pKa$ shift as a function of the salt concentration (Figure S4) results in a non-linear dependence on the logarithm of the salt concentration at very low salt concentrations.
Nevertheless, this dependence becomes again linear when plotted as a function of ionic strength instead of salt concentration.
This subtle difference between the roles of ionic strength and salt concentration was not observed in the experiments of Ref. \cite{ferranddrake20a} because they did not use sufficiently low salt concentrations.
However, it can clearly be observed in our simulations at $\cs\lesssim10^{-5}\,\mathrm{M}$. 

To explain the additional contribution to the $\pKa$ shift, beyond effects accounted for by the Donnan and the SCF mean-field approximation, we use concepts previously introduced in our studies on weak polyelectrolyte hydrogels \cite{landsgesell20b, landsgesell22a, beyer22a}. 
In those studies, we showed that the $\pKa$ shift in two-phase systems can be decomposed into two contributions: $\Delta =  \Delta^\mathrm{PE} + \Delta^\mathrm{Don}$.
The first term, $\Delta^\mathrm{PE}$, expresses the contribution of electrostatic interactions, termed the \emph{polyelectrolyte effect}.
The second term, $\Delta^\mathrm{Don}$, expresses the contribution of the unequal partitioning of \ce{H+} ions, termed the \emph{Donnan effect}.
The magnitude of these effects depends on the parameters of the system.

\begin{figure*}
\begin{subfigure}[t]{0.45\textwidth}
\includegraphics[width=\textwidth]{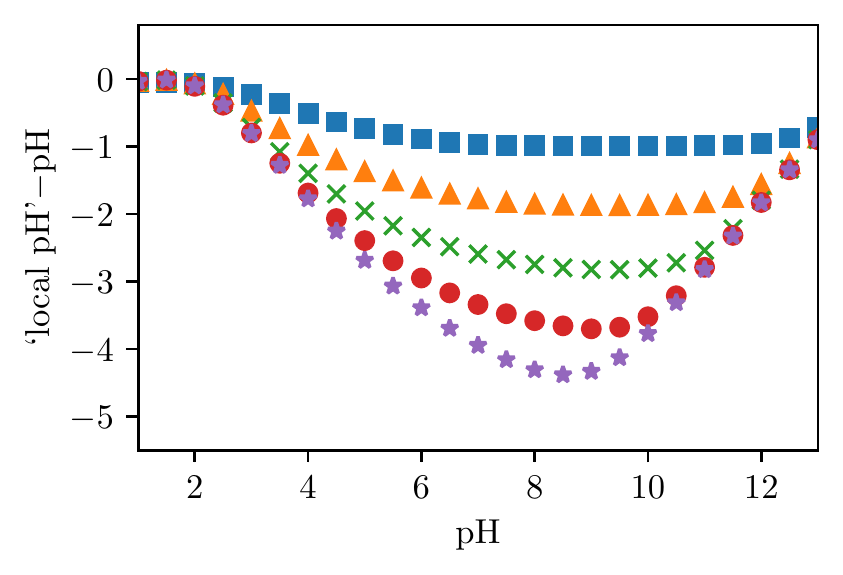}
\caption{\label{fig:delta_pH}}
\end{subfigure}
~
\begin{subfigure}[t]{0.45\textwidth}
\includegraphics[width=\textwidth]{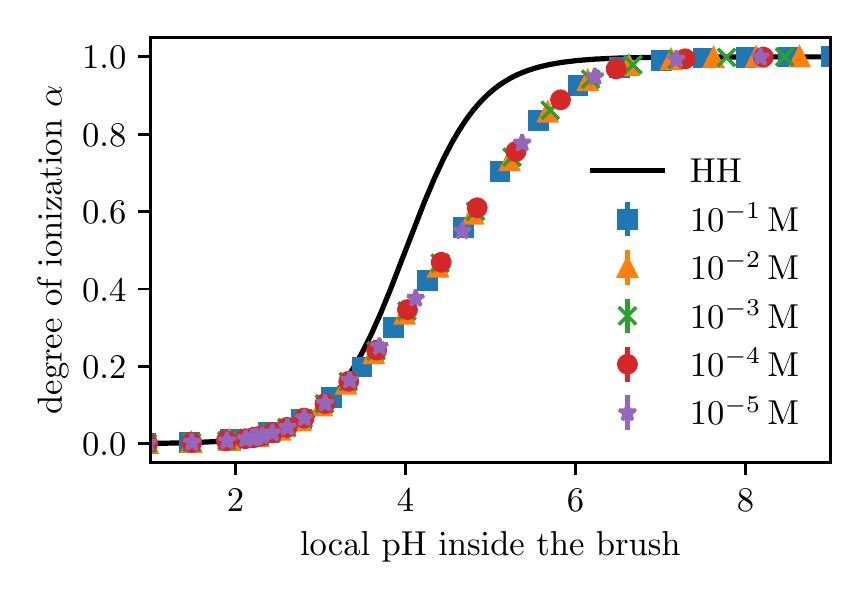}
\caption{\label{fig:alpha_brush_vs_pH_sys}}
\end{subfigure}
\caption{\label{fig:donnan_and_polyelectrolyte}Donnan and polyelectrolyte effect in the brush with a grafting density of $\Gamma=0.079/\mathrm{nm}^2$ as a function of pH at different salt concentrations. 
(a): Difference between the local pH inside the brush and pH in the bulk; 
(b): Degree of ionization of the brush in comparison with the ideal Henderson-Hasselbalch equation (HH).
}
\end{figure*}

To quantify the Donnan effect, we used the density profiles of \ce{H+} ions to determine differences in their concentrations inside and outside the brush.
For convenience, we quantified this difference using the `local pH', defined as
\begin{align}
	\text{local pH} \equiv -\log_{10}\left(\frac{\langle c_\mathrm{H+}\rangle_\mathrm{brush}}{1\,\mathrm{M}}\right) \neq \mathrm{pH}.
\end{align}
The inequality sign in this equation is used to emphasize that the `local pH' is different from the pH in the bulk.
Using this definition, we calculated the local pH inside the brush from the average concentration of \ce{H+} ions up to a distance from the surface of one half of the mean end-to-end distance of the chains at a specific pH and $\cs$.
As evidenced by the monomer density profiles (cf. Figure S6), such a calculation ensures that we average over a range of distances deep inside the brush, where the \ce{H+} concentration is almost constant, thus avoiding artifacts resulting from the brush-solution interface.

\autoref{fig:delta_pH} shows how the difference between the `local pH' inside and outside the brush varies with the pH in the reservoir.
This difference is the quantitative measure of the Donnan effect.
It increases as a function of the pH, concomitantly with the increase in the degree of ionization (cf. \autoref{fig:alpha_brush}).
At $\pH>9$ this difference decreases again although the chains remain fully ionized, because, at a constant salt concentration, an increase in pH necessarily entails an increase in ionic strength.
Furthermore, this difference is larger at lower salt concentrations, in line with \autoref{eq:Donnan-variation}.
By plotting the degree of ionization of the brush as a function of local pH inside the brush, we effectively subtracted the Donnan effect, so that only the polyelectrolyte effect remains. 
In \autoref{fig:alpha_brush_vs_pH_sys}, we show that this polyelectrolyte effect is significant at a low grafting density and accounts for a $\pKa$ shift of approximately 0.5 units of pH.
Figure S8 in the ESI demonstrates that this effect is less significant at a high grafting density.
Regardless of the grafting density, the polyelectrolyte effect remains virtually unchanged at different salt concentrations, leading to a universal behaviour.
To understand this universality, we show plots of $\alpha$ vs. the local ionic strength inside the brush (Figure S10) and of the local ionic strength vs. the local pH (Figure S11) in the ESI.
For the brush at the high grafting density, the degree of ionization is a universal function of the local ionic strength inside the brush.
Furthermore, we find that the local ionic strength inside the brush is uniquely determined by the local pH for values $\geq 3$, i.e. for the relevant pH-range where the brush has acquired an electrical charge.
Combining these insights explains the observed universality in the case of a high grafting density.
In the case of the lower grafting density, we observe a more complex picture.
In particular, for the highest salt concentration (i.e. $0.1\,$M) and for all other salt concentrations at very high pH-values, we the degree of ionization does no longer follow the same $I$-dependence. 
Similarly, a deviation occurs for the ionic strength inside the brush as a function of the local pH.
The fact that this behaviour is correlated with the observed deviation of the swelling (Figure S9b), suggests that the universality emerges due to a cancellation of multiple effects that are related to concomitant changes in the ionization degree and brush swelling.
 We observe that the magnitude of the polyelectrolyte effect matches the difference between the $\pKeff$ calculated from our simulations and from SCF calculations (ca. $0.1$ for the high grafting density and $0.5$ for the low grafting density, cf. \autoref{fig:effective_pK_brush}).
Thus, we have confirmed numerically that the polyelectrolyte effect is the additional contribution to the $\pKa$ shift.

The SCF calculations fully account for the Donnan effect but only approximately account for the polyelectrolyte effect.
Therefore, they quantitatively match simulations at a high grafting density, when the Donnan effect prevails and the polyelectrolyte effect is negligible. 
This agreement weakens at a low grafting density as the polyelectrolyte effect becomes stronger.
These results corroborate our previous findings, demonstrating that SCF can accurately predict the ionization of polyelectrolytes when inter-chain interactions prevail, and that this prediction successively becomes worse as intra-chain interactions get stronger \cite{uhlik14a}.

Our observations of the role of the polyelectrolyte and the Donnan effect could be exploited to infer the grafting density of polyelectrolyte brushes from measured $\pKa$ shifts.
In the ESI, we show that, when the Donnan effect prevails, the grafting density can be approximated by
\begin{align}
	\Gamma\app \approx \frac{2I h_\mathrm{brush}N_\mathrm{A}}{N}\cdot 10^{\Deltadon}
    \label{eq:gamma}
\end{align}
where $N_\mathrm{A}$ is the Avogadro constant.
If we can independently determine the height of the brush, $h_\mathrm{brush}$, the number of monomers per chain, $N$, and the $\pKa$ shift, $\Delta$, we can calculate the apparent grafting density using \autoref{eq:gamma}.
However, the Donnan and polyelectrolyte effect cannot be extracted separately from the experimentally determined $\pKa$ shift $\Delta$.
Therefore, in practice, one has to use $\Delta$ instead of $\Deltadon$ in \autoref{eq:gamma}, assuming that the Donnan effect controls the $\pKa$ shift.
By applying this approach to the densely grafted brush simulated here, we obtain the apparent grafting density $\Gamma\app\approx 0.96/\mathrm{nm}^2$, which is close to the correct value of $\Gamma = 0.79/\mathrm{nm}^2$.
For the less dense brush, we obtain $\Gamma\app\approx 0.25/\mathrm{nm}^2$ which is approximately three times the correct value of $\Gamma =  0.079/\mathrm{nm}^2$.
In general, the apparent grafting density determined using this approach is always an upper bound to the correct value because the polyelectrolyte effect is disregarded.

{\em Conclusion.}
As shown by our simulations, $\pKa$ shifts of weak polyelectrolyte brushes are linear in the logarithm of the ionic strength of the solution, which can be qualitatively explained by the Donnan theory.
However, this logarithmic variation of the $\pKa$ shift does not necessarily imply that the Donnan theory fully describes its physical origin.
This shift is caused by a combination of the Donnan effect and polyelectrolyte effect.
In brushes with a high grafting density, the Donnan effect prevails, and consequently the Donnan theory provides a quantitatively correct description.
In brushes with a lower grafting density, the polyelectrolyte effect significantly contributes to the resulting $\pKa$ shift.
Nevertheless, in both cases, the $\pKa$ varies with the logarithm of the ionic strength.
In short, varying the salt concentration is a general approach to manipulating the effective $\pKa$ of polyelectrolyte brushes, thereby tuning their responsive behaviour in a specific pH range.

\begin{acknowledgments}
D.B. and C.H. and P.K. acknowledge funding by the German Research Foundation (DFG) under the grant
397384169 -- FOR2811. C.H. furthermore thanks the DFG for funding under Project-No 451980436 and 429529433.
P.K. acknowledges funding by the Czech Science Foundation under grant 21-31978J.
We thank Andreas Dahlin for helpful discussions concerning his experimental system.
\end{acknowledgments}

\bibliography{bibtex/icp}

\end{document}


\title{Supplementary Material: Explaining Giant Apparent $\pKa$ Shifts in Weak Polyelectrolyte Brushes}

\author{David Beyer}
\affiliation{Institute for Computational Physics, University of Stuttgart, D-70569 Stuttgart, Germany}

\author{Peter Košovan}
\email{peter.kosovan@natur.cuni.cz}
\affiliation{Department of Physical and Macromolecular Chemistry, Charles University, Prague, Czechia}

\author{Christian Holm}%
\email{holm@icp.uni-stuttgart.de}
\affiliation{Institute for Computational Physics, University of Stuttgart, D-70569 Stuttgart, Germany}

\maketitle

\section{Ideal Donnan Theory}
Within the ideal Donnan theory, we assume that the system can be considered as composed of two compartments (phases): a polymer brush and an aqueous solution containing small ions (\ce{H+}, $\ce{OH-}$, $\ce{Na+}$, $\ce{Cl-}$), termed the bulk.
Additionally, we assume that both the brush and the solution are electroneutral, which implies that the brush confines all of its counterions.
Because both compartments can exchange small ions, an electrochemical equilibrium is established between the two phases: 
\begin{align}
 \mu_i^\mathrm{brush} + z_ie\psi^\mathrm{Don} = \mu_i^\mathrm{bulk}.
 \label{eq:electrochemical_equilibrium}
\end{align}
Within the ideal theory, all interactions are neglected. 
In this case, the Donnan potential $\psi^\mathrm{Don}$ can be obtained as a Lagrange multiplier, originating from the electroneutrality constraint \cite{landsgesell20b}.
For an ideal system, the Donnan equilibrium can be expressed as a relation between the partition coefficient of cations, $\xi_+\equiv c_+^\text{brush}/c_+^\text{bulk}$, and the Donnan potential:
\begin{align}
\xi_+ 
    = \exp\left(-\beta e\psi^\mathrm{Don}\right) 
    = \frac{\alpha\cmon}{2I} + \sqrt{\left(\frac{\alpha\cmon}{2I}\right)^2+1}.
 \label{eq:donnan_partitioning}
\end{align}
Here, $\alpha\cmon$ is the concentration of ionized acid monomers inside the brush and $I$ the ionic strength in the bulk.
Notably, the ionic strength includes all small ionic species $\ce{Na+}$, $\ce{Cl-}$, $\ce{H+}$ and $\ce{OH-}$.
Therefore, the ionic strength is generally different from the concentration of added salt, especially at low salt concentrations and high or low pH values.
The theory thus predicts a difference in the pH-value between the the brush and the bulk solution:
\begin{align}
\begin{split}
    \Delta^\mathrm{Don} 
    &= \log_{10}\left(\exp\left(-\beta e\psi^\mathrm{Don}\right)\right)\\ 
    &\stackrel{\mathrm{ideal}}{=} \log_{10}\left(\frac{\alpha\cmon}{2I} + \sqrt{\left(\frac{\alpha\cmon}{2I}\right)^2+1}\right).
\label{eq:Donnan_effect_brush}
\end{split}
\end{align}  
For a weak polyelectrolyte brush, the degree of ionization, $\alpha$, is not fixed but determined by the pH and Donnan potential inside the brush, related by the Henderson-Hasselbalch equation
\begin{align} 
\alpha = \frac{1}{1+10^{\text{p}K_\text{A}-\text{pH}-\Delta^\mathrm{Don}}}.  
\label{eq:Henderson_Hasselbalch}
\end{align}
\autoref{eq:Henderson_Hasselbalch} and \autoref{eq:Donnan_effect_brush} form a nonlinear system of equations which does not have an analytical solution.

The plot of the degree of ionization of the brush as a function of the pH is shifted by  $\Delta^\mathrm{Don}$, as compared to the same curve in the bulk.  
This shift can be quantified using the effective p$K_\mathrm{A}$, defined as the pH at which the brush is 50\% ionized,
\begin{align}
\mathrm{p}K_\mathrm{A}^\mathrm{eff} 
    \equiv \mathrm{pH}\left(\alpha=\frac{1}{2}\right) 
    \stackrel{\mathrm{ideal}}{=} \mathrm{p}K_\mathrm{A} + \Delta^\mathrm{Don}.
    \label{eq:pkeff}
\end{align}
For a densely grafted brush at $\alpha=0.5$ we have $\alpha\cmon\gg I$ and thus 
\begin{align}
\Delta^\mathrm{Don} &\approx \log_{10}\left(\frac{\cmon}{2I}\right).
\label{eq:Donnan shift}
\end{align}
Because for $\alpha=0.5$ the ionic strength inside the brush is dominated by the counterions confined to the brush, $\cmon$ does not change significantly with $I$ and consequently $\mathrm{p}K_\mathrm{A}^\mathrm{eff}$ decreases approximately linearly with $\log_{10}\left(I/1\,\mathrm{M}\right)$.
In a real interacting system, the $\mathrm{p}K_\mathrm{A}^\mathrm{eff}$ depends not only on the Donnan shift, as suggested by \autoref{eq:pkeff}, but also on additional contributions due to interactions \cite{landsgesell20b}.
In highly charged systems, these additional contributions are dominated by electrostatic interactions, which are neglected in the ideal theory.

If we assume that the $\mathrm{p}K_\mathrm{A}$ shift is dominated by the Donnan term, the we can express the monomer concentration from \autoref{eq:Donnan shift}:
\begin{align}
	\cmon = 2I\cdot 10^{\Delta^\mathrm{Don}}.
\end{align}
In addition, if the degree of polymerization, $N$ is known, then this equation can be used to estimate the grafting density of the brush:
\begin{align}
\Gamma = \frac{2I h_\mathrm{brush}N_\mathrm{A}}{N}\cdot 10^{\Delta^\mathrm{Don}}.
\end{align}
In this formula, $h_\mathrm{brush}$ is the height of the brush and $N_\mathrm{A}$ is the Avogadro constant.

\section{Simulation Model and Methods}

\subsection{Simulation Model}
With our coarse-grained simulation model, we want to represent a grafted weak polyelectrolyte brush.
To model the polymer chains which make up the brush, we make use of the generic Kremer-Grest polymer model with an implicit solvent \cite{grest86a}.
In this model, all particles (monomers and small ions alike) interact via a Weeks-Chandler-Andersen (WCA) potential \cite{weeks71a}:
\begin{align}
V_\text{WCA}(r) = \begin{cases}4\epsilon\left(\left(\frac{\sigma}{r}\right)^{12}-\left(\frac{\sigma}{r}\right)^6\right)+\epsilon &\mbox{if } r\leq2^\frac{1}{6}\sigma\\0 &\mbox{if }r> 2^\frac{1}{6}\sigma.\end{cases}
\end{align}
Here, we set the bead diameter to $\sigma=0.355\,\text{nm}$ and the energy scale to $\epsilon=k_\text{B}T$ with $T= 300\,\text{K}$.
Chemical bonds between adjacent monomers are modelled by the FENE potential \cite{kremer90a}:
\begin{align}
V_\text{FENE}(r) = \begin{cases}-\frac{k\Delta r_\text{max}^2}{2}\ln\left(1-\left(\frac{r-r_0}{\Delta r_\text{max}}\right)^2\right) &\mbox{if } r\leq \Delta r_\text{max}\\ \infty &\mbox{if }r>\Delta r_\text{max}.\end{cases}
\end{align}
In this study we set $\Delta r_\text{max} = 1.5\sigma$ and $k=30k_\text{B}T/\sigma^2$.
Furthermore, charged particles interact via the Coulomb potential,
\begin{align}
V^{ij}_\text{Coulomb}(r) = \frac{\lambda_\text{B}k_\text{B}Tz_iz_j}{r},
\label{eq:Coulomb-Potential}
\end{align}
where we set the Bjerrum length $\lambda_\text{B}=e^2/4\pi\epsilon k_\text{B}T$ to a value of $\lambda_\text{B}=2\sigma=7.1\,\text{\r{A}}$ which corresponds to water at room temperature.

In our simulations, the polyelectrolyte brush of grafting density is represented by an array of $5\times 5=25$ chains with 25 monomers each. 
All monomers are weak acids, characterized by $\mathrm{p}K_\mathrm{A}=4.0$.
This value is close to acrylic acid, which is a typical weak polyacid used in experiments.
To pin the polymer chains to the wall, we fix the first monomer of each chain at a height of $z=\sigma$.
Laterally, the first monomers are fixed on a square lattice, resulting in a uniform grafting density.
The simulation box, corresponding to a slab system, consists of a square cuboid box with lateral dimensions $L_x=L_y=\sqrt{25/\Gamma}$ and height $h=50\sigma$.
To confine the system in the $z$-direction, we make use of fixed walls at $z=0$ and $z=h$ which interact via a WCA potential (with the same parameters as above) with all particles in the simulation.
The simulation box is replicated in all directions using periodic boundary conditions. 
In the $z$-direction, an additional gap is left between the periodic images which is needed to subtract the contributions of the periodic images to the electrostatic interactions \cite{arnold02c}, see below for details.

\subsection{Simulation Method}
To sample different conformations of the system, we make use of Langevin dynamics with $T\approx 300\,\text{K}$, a friction coefficient of $\gamma=1.0$ (in LJ units) and $m=1.0$ (in LJ units).
We integrate the equations of motion using the Velocity-Verlet method \cite{frenkel02b} with a time step of $\Delta t=0.01\,\sigma\left(m/k_\mathrm{B}T\right)^{1/2}$.

In order to simulate the ionization equilibrium of the acidic monomers as well as the exchange of small ions with the bulk, we make use of the G-RxMC method \cite{landsgesell20b}, described in more detail below.
In this method, the simulation box, containing the polyelectrolyte brush and some additional portion of the solution above the brush, is coupled to a reservoir which represents the bulk solution and is characterized by $c_\mathrm{salt}$ and pH.
To efficiently calculate the electrostatic interactions in the slab system, we make use of the P$^3$M method \cite{hockney73a, eastwood80a, hockney88a} with the electrostatic layer correction (ELC) \cite{arnold02c, dejoannis02a, tyagi08a}, which allows us to take into account the interactions with the periodic images only in the $x$- and $y$-directions.
All simulations are carried out using the open-source simulation software ESPResSo \cite{weik19a}.

\subsection{Grand-Reaction Monte-Carlo Method}
In the following we include a detailed description of the employed Grand-Reaction Monte-Carlo method (G-RxMC) \cite{landsgesell20b}.
The dissociation reaction of acidic group, occurring in the polyelectrolyte brush, is schematically described by the equation
\begin{align}
\ce{HA <=> A- + H+}.
\label{eq:HA-ionization}
\end{align}
To circumvent possible bottlenecks in the sampling, we additionally consider the following reactions: 
\begin{align}
\ce{HA <=> &A- + Na+}\\
\ce{HA + Cl- <=> &A-}\\
\ce{HA + OH- <=> &A-}.
\end{align}
As described in Ref. \cite{landsgesell20b}, these reactions are obtained by combining the reaction in \autoref{eq:HA-ionization} with the reactions describing exchange of small ions with the reservoir.
The exchange of small ions with the reservoir is described by the following virtual chemical reactions:
We represent the insertion and deletion of ion pairs by the following set of virtual chemical reactions:
\begin{align}
\emptyset \ce{<=> &H+ + OH-}\\
\emptyset \ce{<=> &Na+ + Cl-}\\
\emptyset \ce{<=> &Na+ + OH-}\\
\emptyset \ce{<=> &H+ + Cl-}.
\label{eq:Insertion-Reactions}
\end{align}
Here, $\emptyset$ denotes the empty set.
To retain the overall electroneutrality of the system, these virtual reactions always insert or delete neutral pairs of ions.
In each reaction step we choose a random reaction and its direction (forward or reverse) with equal probability.
In the forward direction, we insert the particles into the simulation box at random positions.
In the reverse direction, we remove randomly chosen particles from the simulation box or change their identity of particles in accordance with the stoichiometry of the selected reaction.
This proposed new state (n) is accepted according to the following acceptance criterion:
\begin{widetext}
\begin{align}
P^{\mathrm{GRxMC}}_{\textrm{n,o}} = \min \left\{ 1,
\left( \prod_{i} \frac{N_i^0! \left(V \cref\right)^{\nu_i \xi}}{(N_i^0+\nu_i \xi)!}\right)
\exp\left(\beta \left[ \xi \sum_i \nu_i (\mu_i - \muref_i) - \Delta \mathcal{U}_{\textrm{n,o}}\right] \right) \right\}:
\label{eq:sim:g-rxmc}
\end{align}
\end{widetext}
Otherwise the old state (o) is retained. 
In the above acceptance criterion, $V$ denotes the box volume, $\beta = 1/k_{\mathrm{B}}T$ the inverse thermal energy, $\mu_i$ are the chemical potentials, $\nu_i$ the stoichiometric coefficients, $\cref=1\,$M the reference concentration, $\Delta \mathcal{U}_{\textrm{n,o}}$ the change in energy from the old state (o) to the new state (n) and $\xi$ the extent of reaction which is $\xi=1$ for the forward and $\xi=-1$ for the reverse direction of the reaction.
The pH and salt concentration in the reservoir are ultimately determined by the chemical potentials $\mu_i$.
We establish the relation between the imposed pH and $c_\mathrm{salt}$ and the required chemical potentials $\mu_i$ by performing auxiliary simulations in the canonical ensemble of a box filled with ions at different concentrations.
From these simulations we measure the excess chemical potential as a function of the ionic strength using the Widom particle insertion method \cite{widom63a}.
For the imposed values of pH and $c_\mathrm{salt}$ we can then calculate the required chemical potentials from the following set of self-consistent equations:
\begin{align}
\mathrm{pH} &= -\log_{10}\left(\frac{c_{\ce{H+}}}{\cref}\sqrt{\gamma}\right)\\
	\gamma &= f\left(I\left(c_{\ce{H+}},c_{\ce{OH-}},c_{\mathrm{salt}}\right)\right)\\
c_{\ce{OH-}} &= \frac{K_{\ce{H+},\ce{OH-}}\left(\cref\right)^2}{c_{\ce{H+}} \gamma}.
\end{align}
Here, $f(I)$ has to be interpolated using the data obtained from the Widom insertion simulations. 

\subsection{Simulation Protocol}
In order to investigate the influence of the pH value and the salt concentration in the bulk on the behaviour of the brush, we carry out simulations for $\mathrm{pH}\in[1.0,13.0]$ and $c_\mathrm{salt}\in\left[10^{-6}\,\mathrm{M},0.1\,\mathrm{M}\right]$.
To simulate the system, we make use of the following protocol: in the beginning, there are no small ions in the system and all chains are electrically neutral and in a stretched conformation.
To arrive at a random configuration, we run the Langevin thermostat with a random seed for a total of $10^6$ integration steps.
Next, we perform $10^4$ reaction steps in order to add some charged particles to the simulation box. 
This is necessary to tune the P$^3$M solver with ELC in the next step.
After the electrostatics has been added, we minimize the energy using the method of steepest descent.
Then, we run a total of 2500 warmup loops, where in each loop we perform 1000 integration steps and 200 reaction moves, amounting to a total of $5\cdot10^5$ reaction moves.
After this warmup, we retune the electrostatics solver at the new concentration of ions, which is now close to equilibrium.
The production run consists of a total of $2\cdot10^4$ loops. 
In each loop, we perform 1000 integration steps, while we perform 1000 reaction moves every tenth loop.
The observables such as the degree of ionization are sampled every tenth loop.

\subsection{Influence of the Simulation Box Size}
\begin{figure}
\includegraphics[width=.45\textwidth]{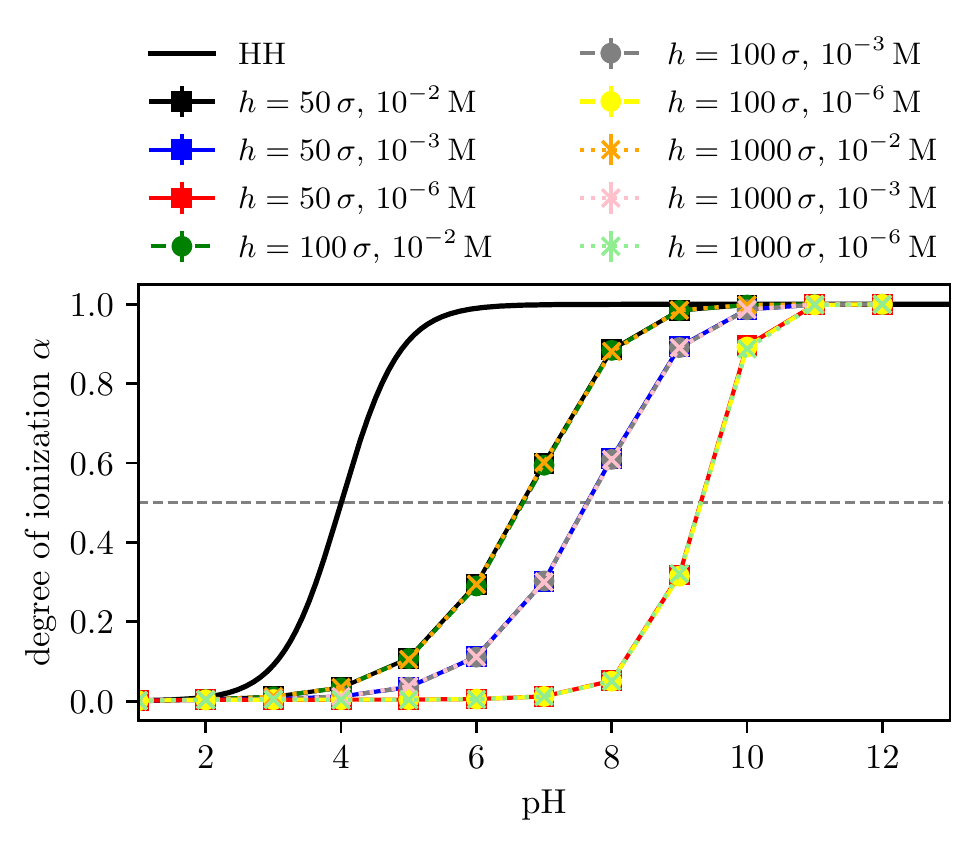}
\caption{\label{fig:alphas_pH_dependent_brush_box_length}Degree of ionization of the brush as obtained for different box sizes and salt concentrations.}
\end{figure}

\begin{figure}
\includegraphics[width=.45\textwidth]{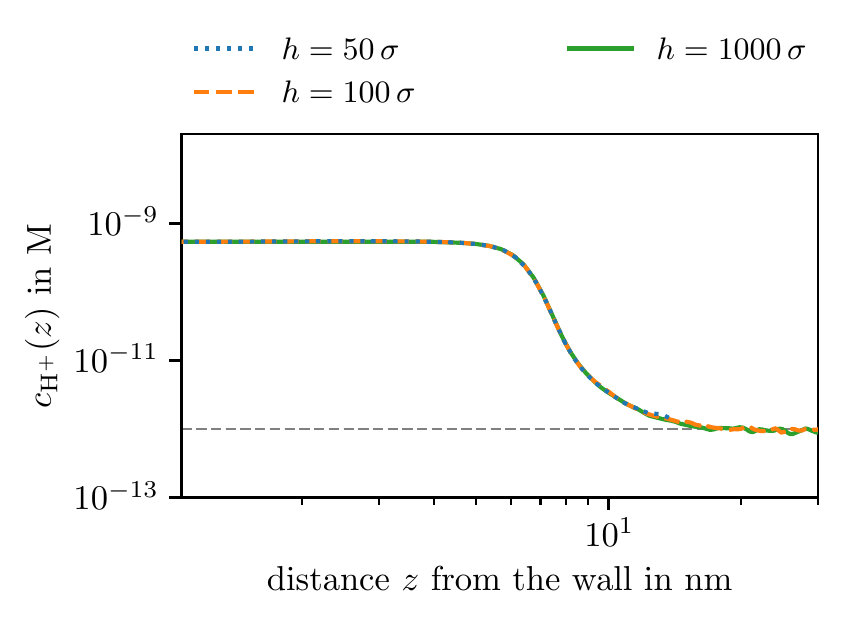}
\includegraphics[width=.45\textwidth]{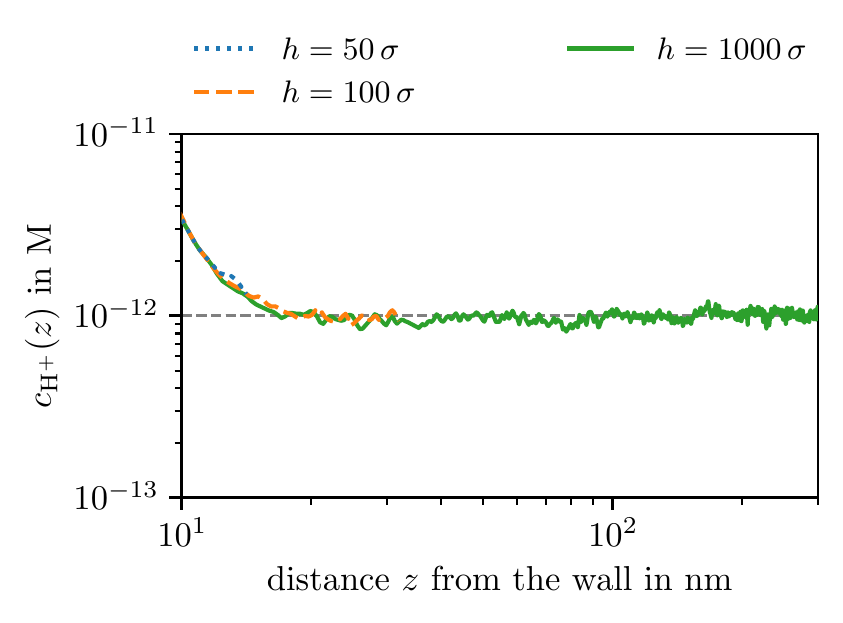}
\caption{\label{fig:density_profile_protons_comparison}Density profiles of \ce{H+} ions for a bulk salt concentration of $c_\mathrm{salt}= 10^{-6}\,\mathrm{M}$ and $\mathrm{pH}=12$, obtained for different simulation box sizes. The dashed line indicates the corresponding \ce{H+} concentrations in the reservoir.
The lower panel shows a zoomed in view of the tail of the distribution.
}
\end{figure}

The simulations presented in the main text were performed for a system with a box height of $h=50\,\sigma=17.75\,\mathrm{nm}$.
We verified that this box size is big enough by performing simulations for a few selected salt concentrations at larger box sizes of $h=100\,\sigma=35.5\,\mathrm{nm}$ and $h=1000\,\sigma=355\,\mathrm{nm}$.
The agreement between the simulation results, shown in \autoref{fig:alphas_pH_dependent_brush_box_length}, justifies a-posteriori the chosen box size.
Irrespective of the simulation box size, we always use the composition of the reservoir as the reference bulk composition.
Otherwise, our calculation would neglect an additional Donnan potential which arises between the simulation box and the reservoir. 
This is evidenced by the proton density profiles in \autoref{fig:density_profile_protons_comparison} which show that these profiles agree within the brush, irrespective of the box size.
For the smallest boxes, the concentration of \ce{H+} at the far end of the simulation box has not yet converged to its bulk value.
For all box sizes used in the current study, this additional contribution is rather small, however, it might become more significant if we had used smaller boxes.
However, being aware of this possible artifact allows us to choose an appropriate box size, which is big enough to be free of artifacts, while simultaneously being small enough in order to avoid excessive computational cost.

\section{Self-consistent field model and method}

The numerical model employed here is based on the Scheutjens-Fleer Self-Consistent Field (SCF) theory \cite{scheutjens79a,fleer93a}.
Its implementation has been described in multiple publications, e.g. in Refs. \cite{wolterink02a,uhlik14a}.
The implementation used here is essentially identical to the one described in full detail in the supporting information of Ref. \cite{uhlik14a}, except that we assume a flat lattice, composed of equally spaced layers, instead of concentric spherical shells.
Therefore, we provide here just a brief description, referring to Ref. \cite{uhlik14a} for the necessary equations and further technical details.

In the Scheutjens-Fleer SCF method \cite{scheutjens79a,fleer97a}, the free energy is expressed as a functional of (i) the volume fraction profiles of the polymer segments, the ions, and the solvent, (ii) the complementary segment potential profiles, and (iii) the Lagrange field (profile) linked to the compressibility condition. 
This free energy functional is optimized on a grid of lattice sites, divided into layers parallel to the surface.
The lattice is subdivided into $z = 1, \ldots, M$ layers.
Properties within each layer are assumed to be uniform while density gradients perpendicular to the surface are considered explicitly.
Furthermore, we assume that all polymer segments solvent molecules and small ions have the same size $z_0 = 0.355\,\mathrm{nm}$, exactly fitting one lattice site which has the volume of $z_0^3$.
We assume additivity of volumes and incompressibility which implies that all volume fractions add up to unity within each lattice layer.
To match the simulation model, we use chains with $N=25$ segments, with the first segment grafted to the surface, i.e. fixed to the first lattice layer.
We assume that the segments are weak acids with $\pKa = 4.0$.
To ensure that our calculations are not too much affected by the finite size of the lattice, we verified that the calculations yield nearly identical results for different system sizes (number of layers): $M\in\{50, 100, 1000\}$.

Within the SCF machinery, we numerically optimise the mean field free energy.
We search for volume fraction profiles of segments, ions, and solvent which minimises the free energy of the whole system and satisfies the incompressibility, electroneutrality and connectivity constraints. 
Simultaneously, we maximise the free energy functional with respect to the segment potentials and the Lagrange field. 
To account for connectivity, we apply the propagator formalism, as described in detail in supporting information of Ref. \cite{uhlik14a}.
The electrostatic interactions are included by solving the one-dimensional Poisson equation for the given density profiles of all charged species. 
The resultant electrostatic potential adds to the mean-field potential and also affects the local chemical potential of charged segments.
The acid-base equilibrium is calculated based on the density profiles of \ce{H+} ions within the Poisson-Boltzmann framework, as described in Refs. \cite{israels94a,israels94b}.
 Excluded volume effects are treated at the Flory-Huggins level with $\chi=0$ (athermal solvent).
The salt concentration and the pH are defined by the boundary conditions as the concentration of each ionic species, including the concentration of \ce{H+} ions, in the outermost lattice layer.
One small ion, either \ce{Na+} or \ce{CL-}, is treated as a neutralizer, \ie, its concentration is adjusted to keep the system electrically neutral.
The equations are solved iteratively until self-consistency between the potentials and density profiles is achieved.

\section{Influence of the Degree of Polymerization}
\begin{figure*}
\begin{subfigure}[t]{0.4999999\textwidth}
\centering
\includegraphics[width=\textwidth]{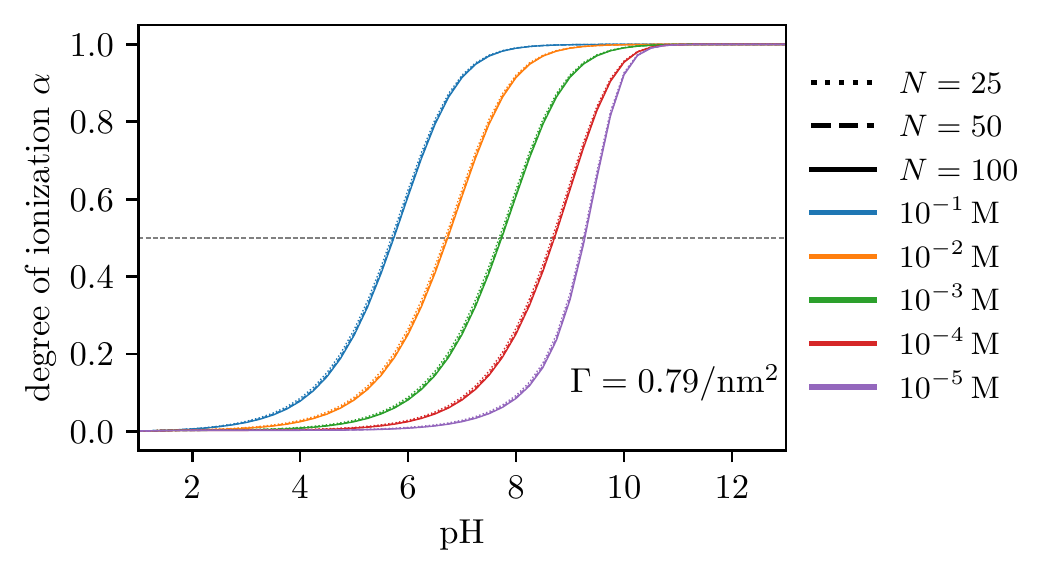}
\caption{\label{fig:fss_high_grafting} $\Gamma=0.79/\mathrm{nm}^2$}
\end{subfigure}%
~
\begin{subfigure}[t]{0.399999\textwidth}
\centering
\includegraphics[width=\textwidth]{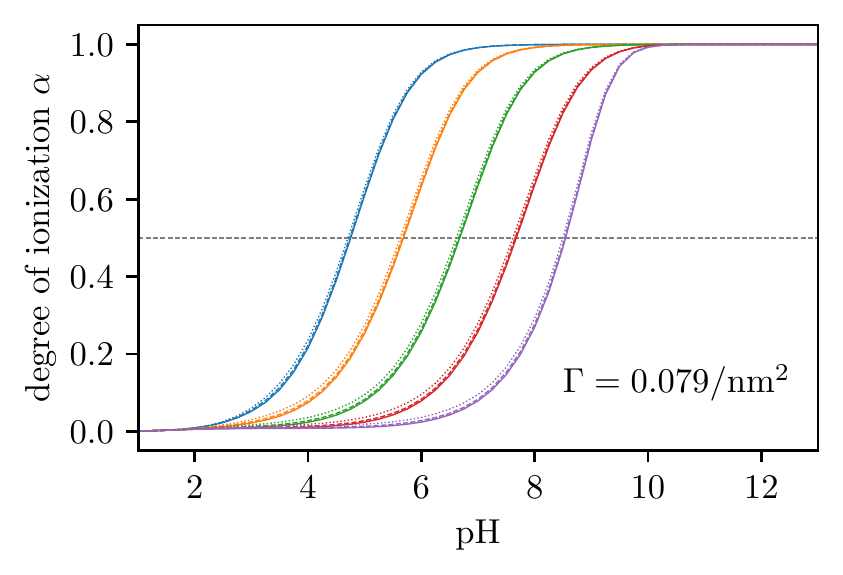}
\caption{\label{fig:fss_low_grafting} $\Gamma=0.079/\mathrm{nm}^2$}
\end{subfigure}%
\caption{\label{fig:fss} Plot of the degree of ionization $\alpha$ vs the pH as obtained using the SCF model for different degrees of polymerization $N\in\left\{25,50,100\right\}$.
The results for different $N$ are shown as dotted, dashed and full lines respectively. 
The two subfigures show analogous results for different grafting densities of the brush.
}
\end{figure*}

In addition to the chosen box size (which was addressed above), we also investigate the influence of the brush height, i.e. the degree of polymerization $N$ of the chains, on the result.
Generally, one would expect that the influence of the explicit brush-solution interface decreases as $N$ is increased.
While such a finite-size scaling study would be prohibitively expensive on the particle-based level, it can be carried out rather easily on the SCF level.
The in general good agreement between SCF and particle-based simulations justifies this approach.
To probe the influence of the degree of polymerization, we carried out SCF calculations for brushes with $N\in\left\{25,50,100\right\}$.
\autoref{fig:fss} shows the resulting ionization curves as a function of the pH.
The plot shows that there is a miniscule shift to the right, i.e. the ionization is slightly more suppressed, as $N$ is increased. 
This is expected, because the relative importance of the interface, where the local pH is higher than inside the brush, diminishes as the chains become longer.
Furthermore, we observe that the shift seems to saturate with increasing $N$, i.e. the shift when going from 25 to 50 is larger than when going from 50 to 100.
This observation thus justifies the chosen value of $N=25$ for the particle-based simulations.

\section{Additional Figures}
\begin{figure}
\includegraphics[width=.45\textwidth]{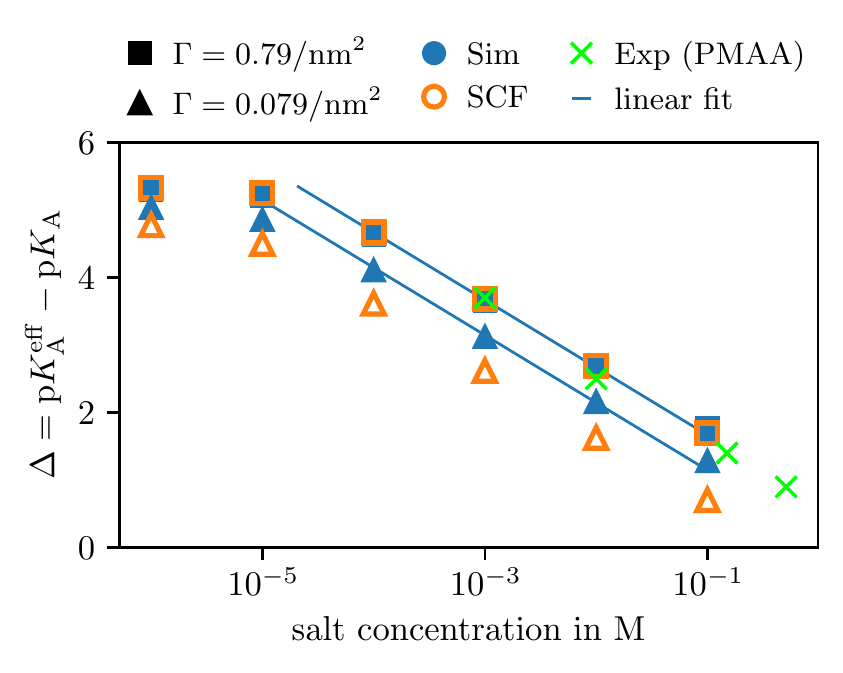}
\caption{\label{fig:effective_pK_brush_vs_salt}
The $\pKa$ shift of brushes with different grafting densities as a function of salt concentration. 
Solid symbols represent data from particle-based simulations, whereas empty symbols represent SCF results. 
}
\end{figure}

\begin{figure}
\includegraphics[width=.45\textwidth]{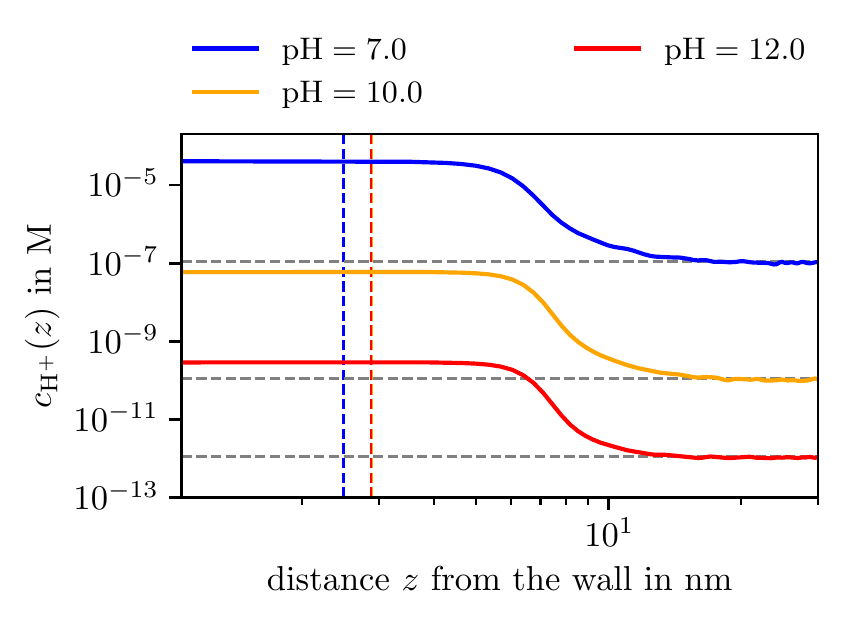}
	\caption{\label{fig:density_profile_protons}Proton density profiles for a bulk salt concentration of $c_\mathrm{salt}= 10^{-2}\,\mathrm{M}$ and various pH-values. The vertical dashed lines indicate the value of $\langle R_\mathrm{e}\rangle/2$ (for $\mathrm{pH}=10$ and $\mathrm{pH}=12$, the values cannot be distinguished visually). The horizontal dashed lines indicate the corresponding \ce{H+} concentrations in the reservoir.
     }
\end{figure}

\begin{figure*}
\begin{subfigure}[t]{0.45\textwidth}
\centering
\includegraphics[width=\textwidth]{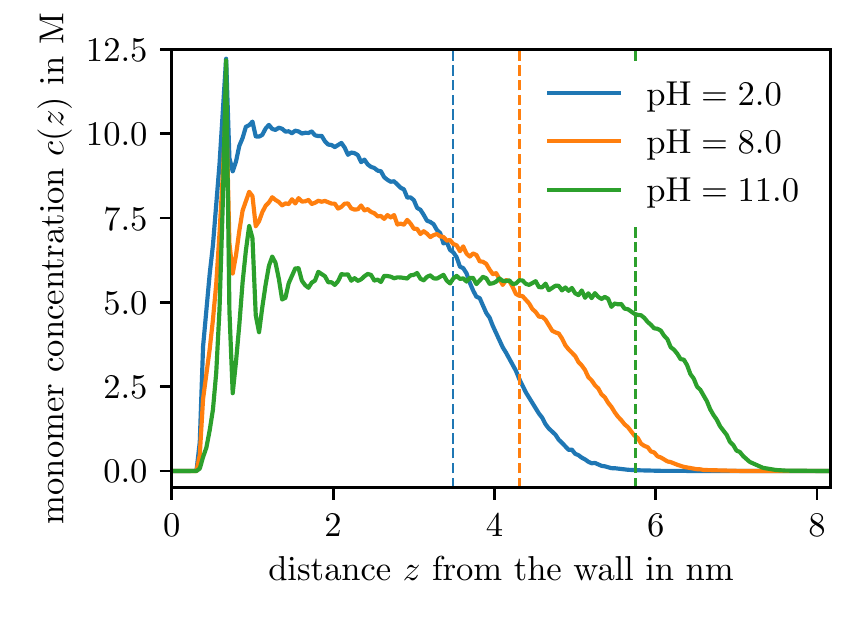}
\caption{\label{fig:monomer_profile_brush_high_grafting} $\Gamma=0.79/\mathrm{nm}^2$}
\end{subfigure}%
~
\begin{subfigure}[t]{0.45\textwidth}
\centering
\includegraphics[width=\textwidth]{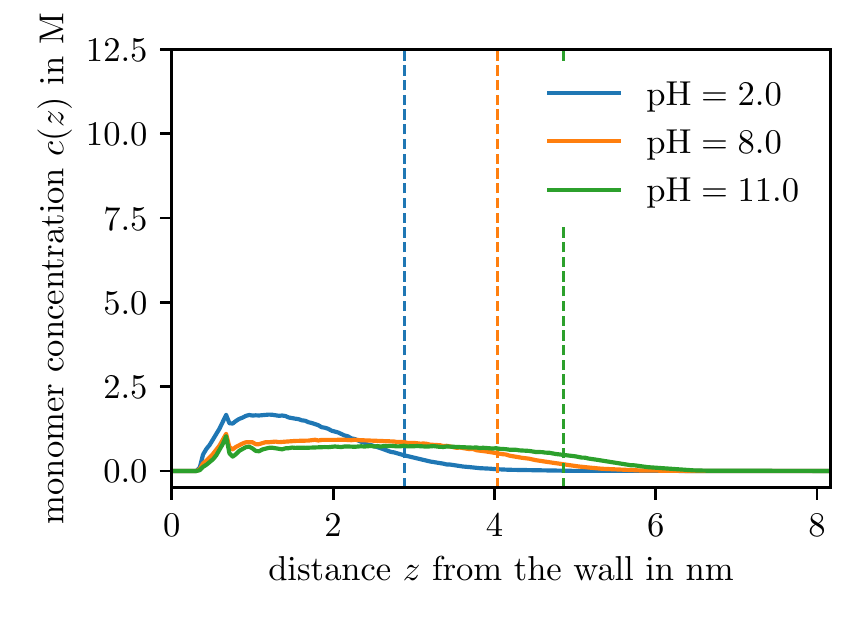}
\caption{\label{fig:monomer_profile_brush_low_grafting} $\Gamma=0.079/\mathrm{nm}^2$}
\end{subfigure}%
\caption{\label{fig:monomer_profile_brush_comparison}Monomer concentration profiles of the brush as a function of the distance $z$ from the wall for $c_\mathrm{salt}=10^{-4}\,\mathrm{M}$ and different values of the pH in the reservoir.
The dashed vertical lines indicate the mean values of the end-to-end distance $\langle R_\mathrm{e}\rangle$.
The two subfigures show analogous results for different grafting densities of the brush.
}
\end{figure*}

\begin{figure*}
\begin{subfigure}[t]{0.45\textwidth}
\centering
\includegraphics[width=\textwidth]{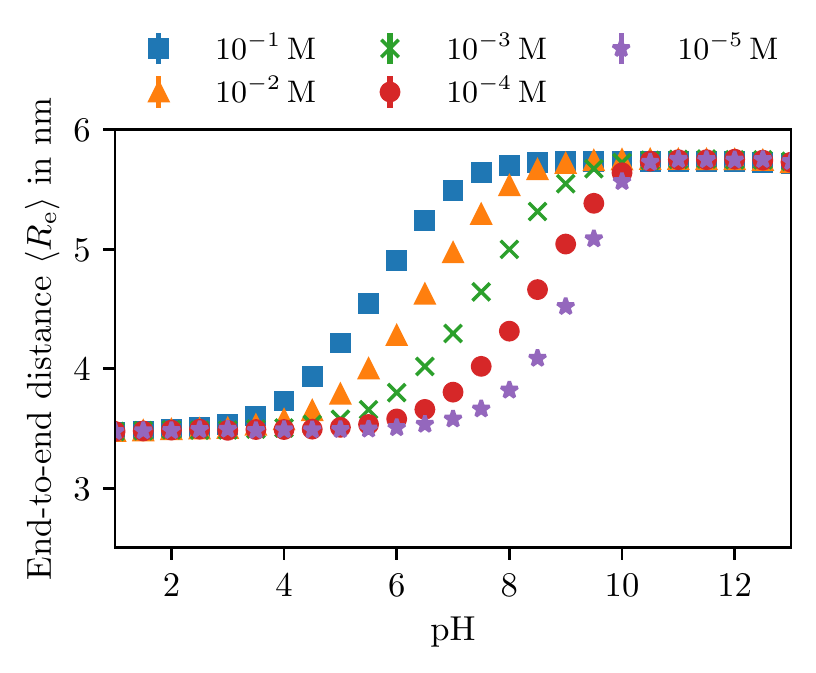}
\caption{\label{fig:brush_height_comparison_high_grafting} $\Gamma=0.79/\mathrm{nm}^2$}
\end{subfigure}%
~
\begin{subfigure}[t]{0.45\textwidth}
\centering
\includegraphics[width=\textwidth]{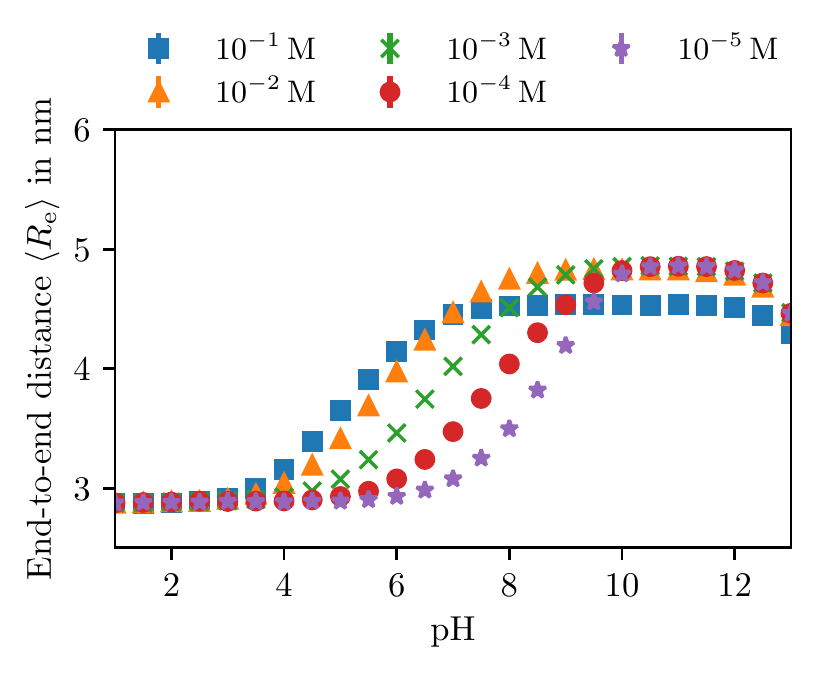}
\caption{\label{fig:brush_height_comparison_low_grafting} $\Gamma=0.079/\mathrm{nm}^2$}
\end{subfigure}%
\caption{\label{fig:brush_height_comparison} Plot of the pH-dependent mean end-to-end distance $\langle R_\mathrm{e}\rangle$ of the chains for the brush.
The two subfigures show analogous results for different grafting densities of the brush.
}
\end{figure*}

\begin{figure*}
\begin{subfigure}[t]{0.45\textwidth}
\centering
\includegraphics[width=\textwidth]{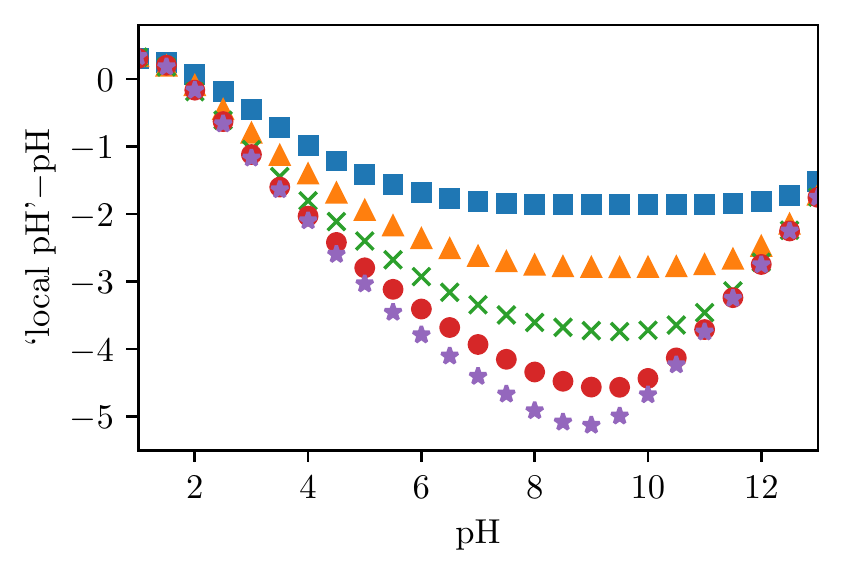}
\caption{\label{fig:degree_of_ionization_brush_vs_pH_sys_comparison_high}}
\end{subfigure}%
~
\begin{subfigure}[t]{0.45\textwidth}
\centering
\includegraphics[width=\textwidth]{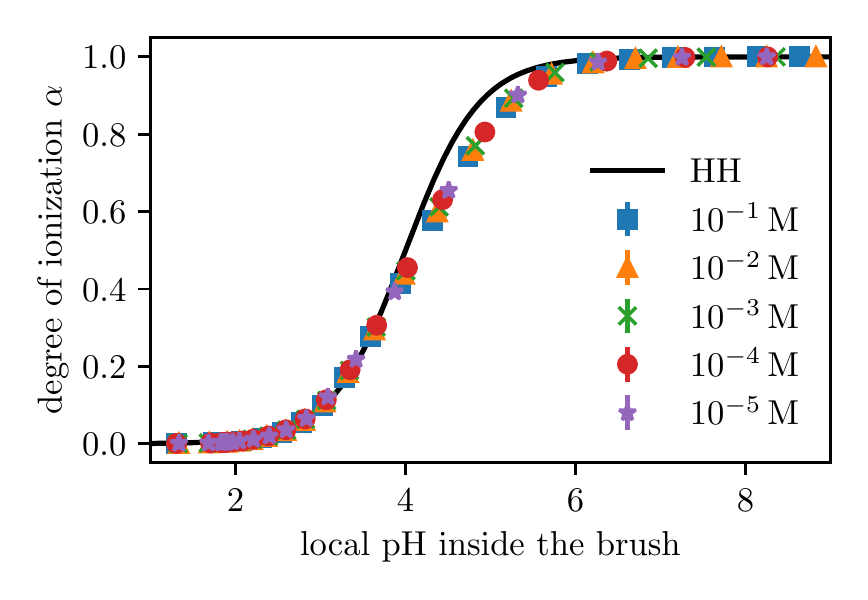}
\caption{\label{fig:degree_of_ionization_brush_vs_pH_brush_comparison_high}}
\end{subfigure}%

\caption{\label{fig:degree_of_ionization_brush_vs_pH_sys_comparison}
Donnan and polyelectrolyte effect in the brush with a grafting density of $\Gamma=0.79/\mathrm{nm}^2$ as a function of pH at different salt concentrations. 
(a): Difference between the local pH inside the brush and pH in the bulk; 
(b): Degree of ionization of the brush in comparison with the ideal Henderson-Hasselbalch equation (HH).
}
\end{figure*}

\begin{figure*}
\begin{subfigure}[t]{0.45\textwidth}
\centering
\includegraphics[width=\textwidth]{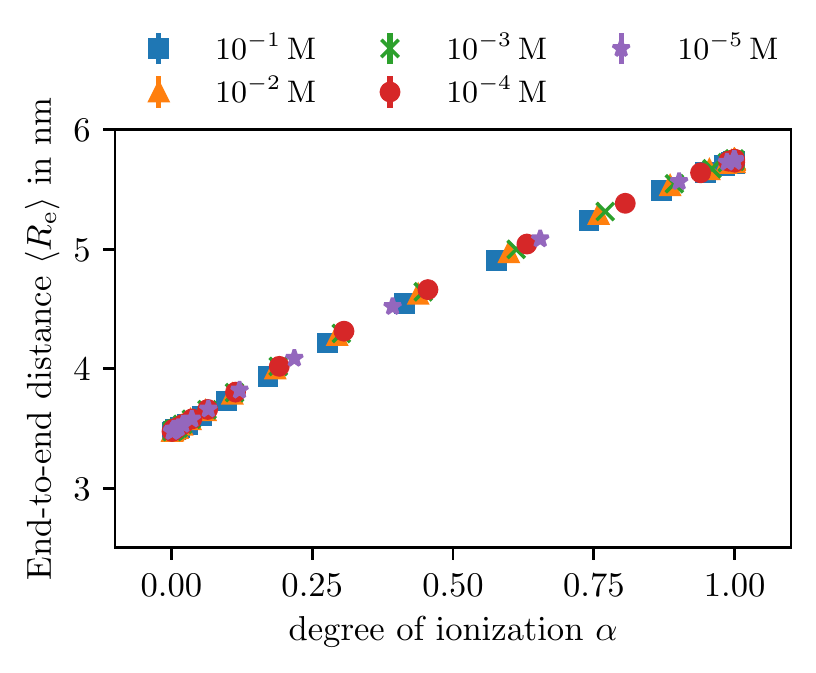}
\caption{\label{fig:brush_height_alpha_comparison_high_grafting} $\Gamma=0.79/\mathrm{nm}^2$}
\end{subfigure}%
~
\begin{subfigure}[t]{0.45\textwidth}
\centering
\includegraphics[width=\textwidth]{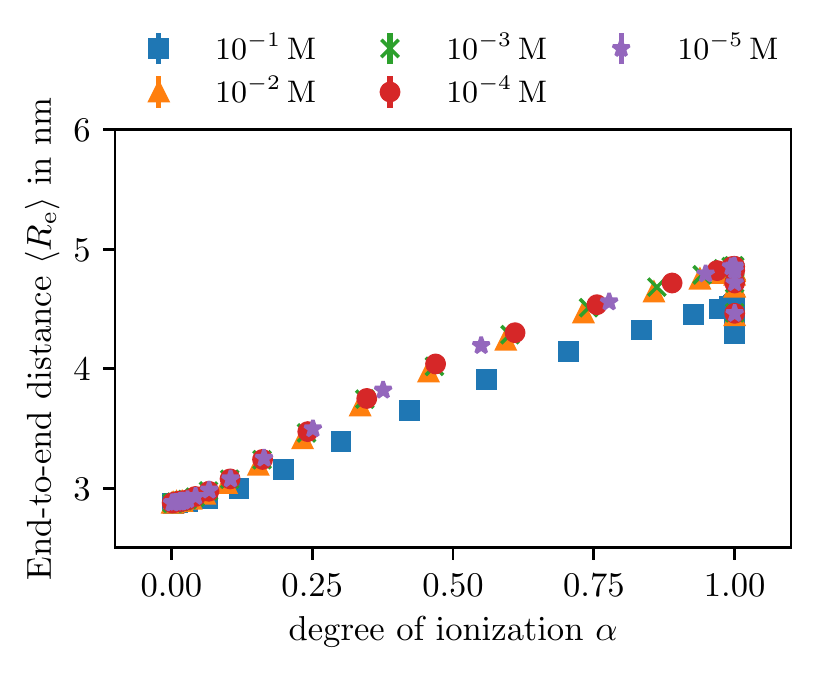}
\caption{\label{fig:brush_height_alpha_comparison_low_grafting} $\Gamma=0.079/\mathrm{nm}^2$}
\end{subfigure}%
\caption{\label{fig:brush_height_comparison_alpha} Plot of the $\alpha$-dependent mean end-to-end distance $\langle R_\mathrm{e}\rangle$ of the chains for the brush.
The two subfigures show analogous results for different grafting densities of the brush.
}
\end{figure*}

\begin{figure*}
\begin{subfigure}[t]{0.45\textwidth}
\centering
\includegraphics[width=\textwidth]{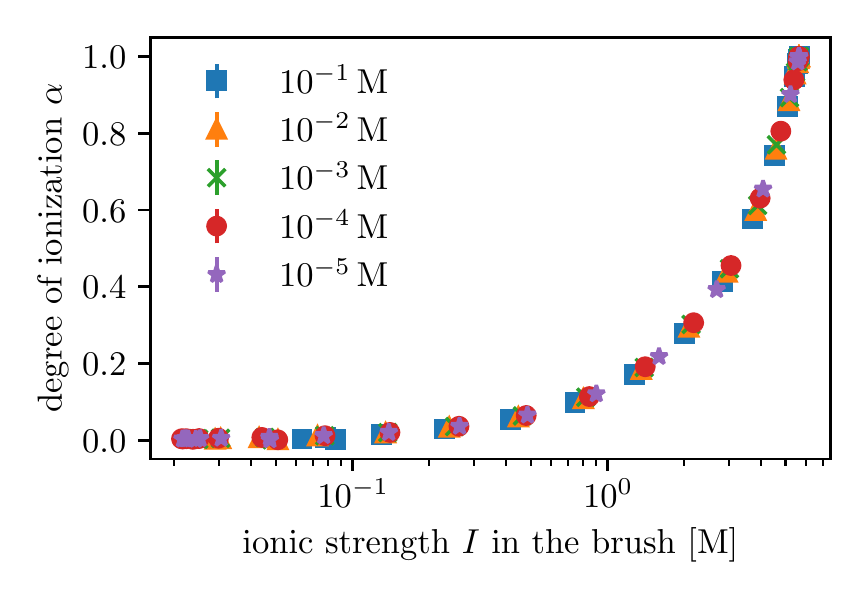}
\caption{\label{fig:alpha_vs_ionic_strength_high_grafting} $\Gamma=0.79/\mathrm{nm}^2$}
\end{subfigure}%
~
\begin{subfigure}[t]{0.45\textwidth}
\centering
\includegraphics[width=\textwidth]{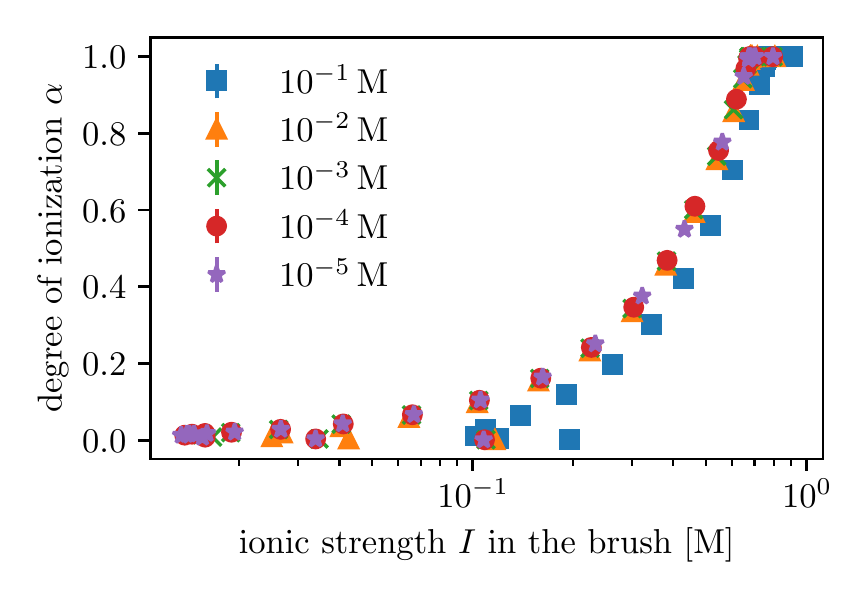}
\caption{\label{fig:alpha_vs_ionic_strength_low_grafting} $\Gamma=0.079/\mathrm{nm}^2$}
\end{subfigure}%
\caption{\label{fig:alpha_vs_ionic_strength} Plot of the degree of ionization $\alpha$ as a unction of the ionic strength inside the brush.
The two subfigures show analogous results for different grafting densities of the brush.
}
\end{figure*}

\begin{figure*}
\begin{subfigure}[t]{0.45\textwidth}
\centering
\includegraphics[width=\textwidth]{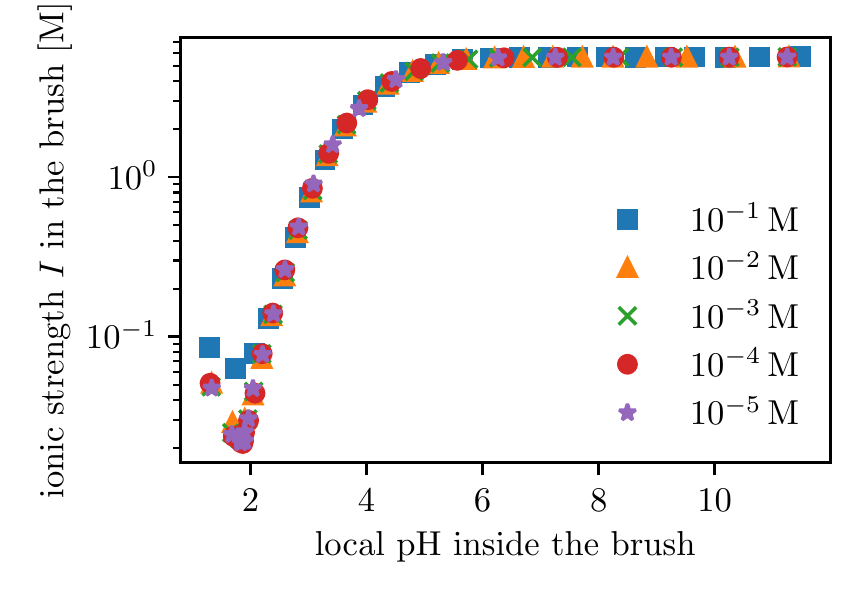}
\caption{\label{fig:ionic_strength_vs_pH_sys_high_grafting} $\Gamma=0.79/\mathrm{nm}^2$}
\end{subfigure}%
~
\begin{subfigure}[t]{0.45\textwidth}
\centering
\includegraphics[width=\textwidth]{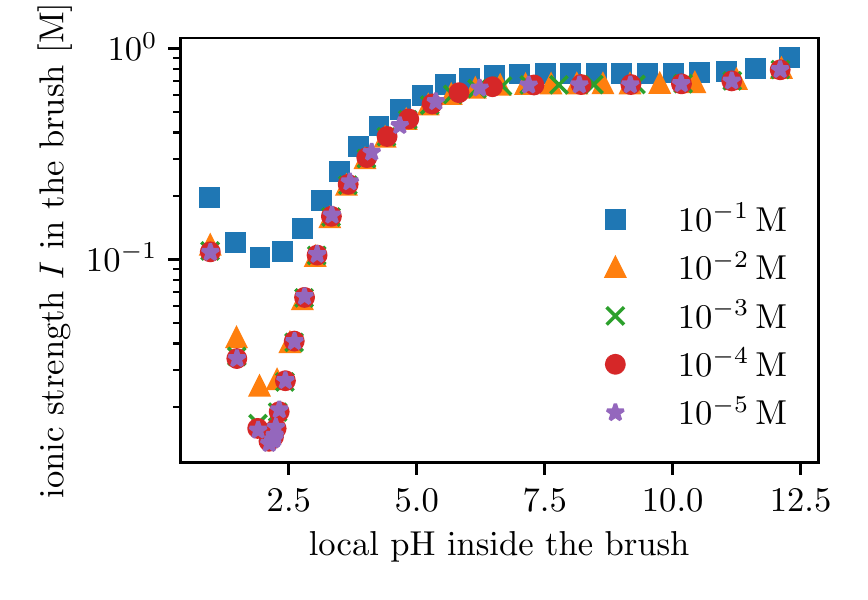}
\caption{\label{fig:ionic_strength_vs_pH_sys_low_grafting} $\Gamma=0.079/\mathrm{nm}^2$}
\end{subfigure}%
\caption{\label{fig:ionic_strength_vs_pH_sys} Plot of the ionic strength inside the brush as a function of the local pH inside the brush.
The two subfigures show analogous results for different grafting densities of the brush.
}
\end{figure*}

\begin{figure*}
\begin{subfigure}[t]{0.45\textwidth}
\centering
\includegraphics[width=\textwidth]{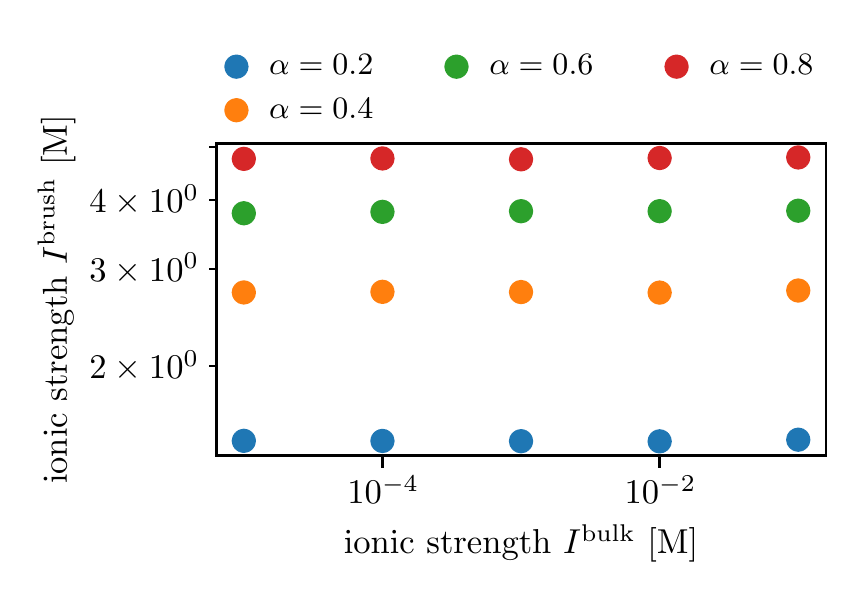}
\caption{\label{fig:ionic_strength_brush_vs_ionic_strength_sys_high_grafting} $\Gamma=0.79/\mathrm{nm}^2$}
\end{subfigure}%
~
\begin{subfigure}[t]{0.45\textwidth}
\centering
\includegraphics[width=\textwidth]{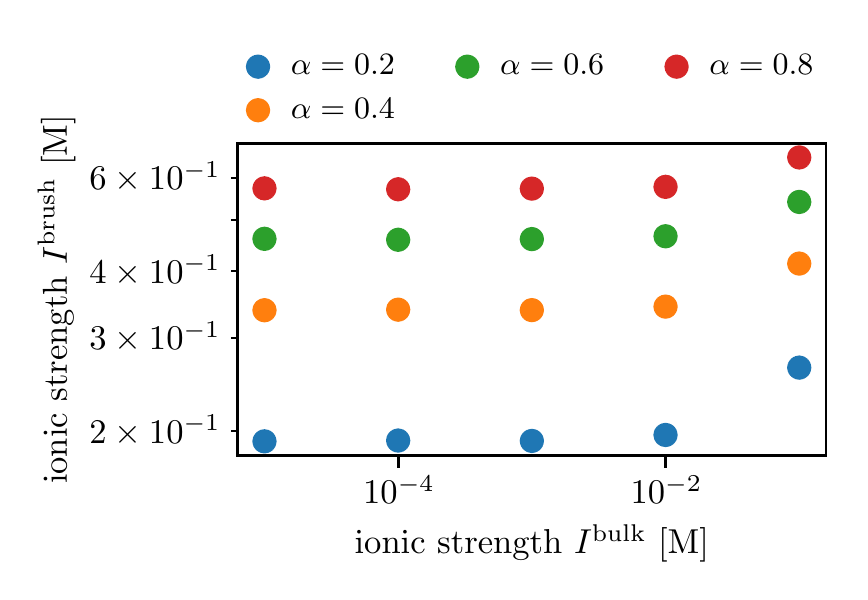}
\caption{\label{fig:ionic_strength_brush_vs_ionic_strength_sys_low_grafting} $\Gamma=0.079/\mathrm{nm}^2$}
\end{subfigure}%
\caption{\label{fig:ionic_strength_brush_vs_ionic_strength_sys} Plot of the ionic strength inside the brush as a function of the ionic strength in the bulk solution at various fixed values of the degree of ionization $\alpha$.
The two subfigures show analogous results for different grafting densities of the brush.
}
\end{figure*}

\begin{itemize}
\item \autoref{fig:effective_pK_brush_vs_salt} shows the $\pKa$ shift as a function of the salt concentration rather than the ionic strength (\autoref{fig:effective_pK_brush}). At very low salt concentrations, this representation leads to a nonlinear behaviour.
\item Selected concentration profiles for the protons as obtained from the particle-based simulations, calculated using a kernel density estimation, are shown in \autoref{fig:density_profile_protons}. The profiles, obtained from the simulations with a box size of $h=1000\,\sigma$, display a gradual decay rather than a sharp transition that is one of the assumptions of the ideal Donnan theory. Far away from the brush, the concentration approaches the bulk/reservoir value. For values of $z$ smaller than $\langle R_\mathrm{e}\rangle/2$ the profiles are effectively constant.
This distance was used to define the region "inside of the brush".
\item In \autoref{fig:monomer_profile_brush_comparison} we show selected monomer concentration profiles of the brush at different pH-values and grafting densities. 
The distribution gets broader as the pH is increased, because the brush becomes ionized and swells due to the osmotic pressure of the counterions and the electrostatic repulsion.
Another interesting feature one can observe is the appearance of sharply defined maxima and minima in the distribution function near the wall as the swelling increases. 
These appear because the chains become more stretched and near the wall, where the chain ends are fixed, this increases the ordering.
\item \autoref{fig:brush_height_comparison} displays the pH-dependent mean end-to-end distance of the brush, quantifying the swelling behaviour of the brush.  
\item \autoref{fig:degree_of_ionization_brush_vs_pH_sys_comparison} shows the Donnan and polyelectrolyte effect in the brush with a grafting density of $\Gamma=0.79/\mathrm{nm}^2$ as a function of pH at different salt concentrations (this is analogous to \autoref{fig:donnan_and_polyelectrolyte} in the main text).
\item \autoref{fig:brush_height_comparison_alpha} shows the mean end-to-end distance of the brush as a function of the degree of ionization of the brush.
For the brush at the higher grafting density, we observe that the swelling curves for the different bulk salt concentrations collapse onto a universal swelling curve.
This behaviour suggests that the densely grafted brush can be considered to be in the osmotic brush regime for all considered salt concentrations.
In contrast, for the brush at the lower grafting density, we observe that while the data collapses onto a universal curve for the lower salt concentrations, the brush swells less for the highest investigated salt concentration ($0.1\,$M).
Furthermore, for all curves we observe a deswelling at very high pH-values, which is related to the concomitant increase of the bulk ionic strength.
These observations imply that the brush transitions from the osmotic regime to the salted regime at a high enough salt concentration.
\item \autoref{fig:alpha_vs_ionic_strength} shows the degree of ionization $\alpha$ of the brush as a function of the ionic strength inside the brush (including the charged monomers).
\item \autoref{fig:ionic_strength_vs_pH_sys} shows the ionic strength inside the brush (including the charged monomers) as a function of the local pH inside the brush.
\item \autoref{fig:ionic_strength_brush_vs_ionic_strength_sys} shows the ionic strength inside the brush as a function of the ionic strength in the bulk solution at various fixed values of the degree of ionization $\alpha$.
\end{itemize}

\bibliography{bibtex/icp}